\documentclass{PoS}
\usepackage{amsmath,amssymb}
\title{What if? On the interplay between Serendipity, 
Intuition and Conjecture }

\ShortTitle{What if?}

\author{\speaker{Benjam\'\i n Grinstein}\thanks{The author wishes to
    thank the theory group at CERN for their hospitality. Work supported in part by
  U.S. Department of Energy under Contract No. DE-FG03-97ER40546. }\\
        UCSD and CERN\\
        E-mail: \email{bgrinstein@ucsd.edu}}

      \abstract{While the Standard Model is in good shape, there are
        many reasons to believe it is incomplete. There are high
        expectations that the LHC will shed light on some well studied
        possibilities, like technicolor and supersymmetry. Emboldened
        by this optimism, we consider some non-mainstream ideas that
        if established would change dramatically the way we view the
        world. }

\FullConference{Flavor Physics and CP Violation - FPCP 2010\\
		May 25-29, 2010\\
		Turin, Italy}

\begin{document}

\section{Introduction: Now and Then}
This may seem like an odd title for a talk at this conference, a
Summary Talk no less! So before launching into it let me justify
this. First, there seems little point in giving an actual
summary. There were brilliant progress reports over four days of talks
and several summary talks, including the excellent next-to-last talk
by Hassan Jawahery.  On the theory side, in addition to Vicenzo
Cirigliano's status review of charged lepton flavor violation and
Thomas Schwetz-Mangold's report on neutrino physics, we had two talks
that completely encapsulated the output of flavor physics and CPV(iolation) (in
the quark sector), namely, Enrico Lunghi's Lessons from CKM Studies
and Zoltan Ligeti's talk on Physics Reach of Future Flavor
Experiments.  In turn, their reports and summaries rely on much
other theory work reported here, {\it e.g.},  Jure
Zupan on CKM angles, Guido Bell on theoretical issues in hadronic
decays and Gilad Perez on D decays. Two of my favorite subjects
(favorite since I surely owe my tenure job to them) were magnificently
covered by old friends Mikolaj Misiak, on radiative B
decays, and Thomas Mannel, on determination of $V_{cb}$ and
$V_{ub}$. The interpretation of experimental results is becoming clearer
with ever improving Lattice studies, as reported by Jack Laiho and
another old friend, Junko Shigemitsu.  

Going back to the title\dots given all these talks, and particularly
the comprehensive summaries, I hope you will agree with me that it is
unnecessarily repetitious and exceedingly boring to present a grand
final ``Summary and Conclusions.''  So I thought it'd be better and
more fun to talk about the big picture. And since the LHC is starting
to collect data, the time is ripe for that. However, to paraphrase
Frank Wilczek, {\it an unfortunate byproduct of the delays at the LHC
  and the slow pace of other experiments is that general talks about
  the grandeur of high energy physics are getting
  stale.}\cite{Wilczek:2010br} While previous installments of FPCP
have had few, if any, such grand talks, surely most in the audience
had heard some elsewhere. So I would like to take a different track.
But before doing so, let me summarize Wilczek's Litany, just to make
sure we all agree on the grandeur of our field and its exciting
future.  On the one hand the standard model of strong and electroweak
interactions combined with the CKM framework is fantastically
successful. Excellent agreement between theory and experiment severely
constraints possible extensions of the model. On the other hand, there
are many shortcomings of the model, among them, lack of understanding
of the smallness and nature of neutrino masses, no accounting for dark
matter and dark energy, no explanation of the smallness of the P and T
violating $\theta$ parameter, and, of particular interest to FPCPers,
no explanation for the triplication of families and a lack of
fundamental principles to constrain the numerous masses and mixing
angles.\footnote{In my talk I let Jim Cronin present these last two
  shortcomings, by showing an excerpt from his video-recorded talk in
  \cite{cronin}.}  But as I said, I want to discuss something else. I
am interested in the possibility of paradigm shifting discoveries. I
will give you my views on the possibility of such discoveries at the
LHC towards the end of the talk. But for now let us recall that our
field, both FP and CP in FPCP, resulted from such discoveries.

In 1935 Hideki Yukawa published his theory of mesons, which explained
the interaction between protons and neutrons.  Muons were discovered
by Carl D. Anderson, and student Seth Neddermeyer, in 1936, while they
studied cosmic radiation using a cloud chamber. Incidentally, on that
same year Anderson received the Nobel prize for the discovery of the
positron.  In 1947, that's 11 years later, the name changed to
mu-meson to differentiate it from the many other hadronic mesons being
discovered.  The mesotron was initially thought to be the pi-meson of
Yukawa. I did not have the time to verify this, but I believe it is
true, that Yukawa's work was an important motivation for Anderson's
research.  The discovery of the muon was a complete surprise.  ``Who
ordered that?'' quipped I.I. Rabi. Serendipity certainly played
a role in the birth of flavor physics.  The discovery was motivated by
a theory which was irrelevant but was not quite incorrect.

This is what the title of the talk stands for: conjecture or
speculation, right or wrong, can motivate a good experiment.
Intuition is needed to follow the right path.  Serendipity cannot
hurt.  As a theorist I find some solace in the possibility that a
Theory of Nothing,\footnote {By which I mean a model of nature that
  happens to be wrong because it does not describe reality; see also \cite{Weinberg:1992nd}.} combined
with Good Luck and a Good Experiment can lead to a Discovery. In the
above example the theory was not really wrong, but was irrelevant to
the discovery.  Correct theoretical ideas can motivate an experiment,
of course. Anderson's (1932) discovery of the positron was a direct
response to Dirac's (1928) theory. What I am after is that sometimes a
theory is just crazy enough that may push us to think about tests we
have not carried out. And I am also saying that in some sense this is
how our field of FPCP was born.

Serendipity played a role in the discovery of CP violation. That CP is
not respected by nature was a surprise to the discoverers. There was
no theory at the time, so there was no level of CP violation to aim
for. For example, Sakharov's conditions for baryogenesis, which
include T-reversal violation, were only published in 1967, four years
later than Cronin, Fitch and Turlay submitted their proposal.  The
experiment was designed to measure better something (regeneration) for
which there was a good knowledge base and in passing to improve limits
on other impossibilities (like the branching fraction on
$K_L\to\pi\pi$).  I find remarkable Cronin's statement\cite{cronin}
that he does not know how they came about with the wrong estimate for
sensitivity to CP violation in the proposal. This may be evidence that
the experimenters did not take that aspect of the proposal as
central. Testing for CP non-invariance was clearly worth doing, but
not enough to put much effort into estimating the sensitivity {\it
  before} the project was approved.  We can draw many other examples
from history, not necessarily from FPCP. I will mention only one
more. The proposal of Grand-Unified theories gave rise to a race to
detect proton decay. As it turns out, as you all know, the theory is
wrong. To be precise, the experiments were designed to test what we
now call ``minimal non-SUSY SH5.'' They managed to rule it out, well
before electroweak precision experiments demonstrated that the
couplings do not quite unify.  But the experiments, it was soon
discovered, make for wonderful neutrino telescopes, and made a number
of remarkable discoveries, among them, the detection of neutrinos from
supernova 1987a, and the oscillation of atmospheric neutrinos.

So I propose to you to take a sampling of non-mainstream, almost
surely wrong, ideas. Some have well motivated theory. Some don't. The
main criterion is that confirmation of any would result in a paradigm
shift.\footnote{From Wikipedia: ``Paradigm shift \dots is the term
  first coined by Thomas Kuhn in his influential book The Structure of
  Scientific Revolutions (1962) to describe a change in basic
  assumptions within the ruling theory of science. ''}  Because of
time limitations I just selected a few examples; this is not intended
to be a comprehensive presentation. Also, I do not consider some of
our favorite extensions of the SM to be paradigm shifting: in models
like the MSSM, horizontal gauged symmetry, technicolor, unphysics,
Little Higgs, etc, the basic tenets remain intact.  Sure, these models
require additional fields and interactions and it would be exciting
and interesting to find out that nature is like any one of these. But
the basic principles of modern particle physics, {\it e.g.,} that
physics is described by a local relativistic quantum field theory
which yields an analytic, causal, $S$-matrix, are
unchanged. Extra-dimensions is, in my view, paradigm-shifting, and so
is string theory. The examples I have chosen to discuss below are,
however, less main-stream (or really ``out there''). 

\section{Violation of CPT and/or QM}
The CPT Theorem tells us that in a quantum mechanical model of local
fields, with dynamics dictated by a hermitian Hamiltonian, one can
define a discrete operation CPT that is a symmetry of the model, even
when the individual operations $C$ and $P$ may not be well defined.
Hence CPT is a symmetry of the SM. But observation of violation of CPT
would not only clearly require to amend the SM, but would indicate a
violent departure of our current paradigm that insists on describing
the world on the basis of local Quantum Field Theory.  String theory
and loop Quantum Gravity (QG) are but two examples of such violent
departures, formulated, in terms of non-local, non-field
theories. Generally, theories of QG are expected to violate CPT. The
argument is simple, that BHs cannot carry non-gauged conserved
numbers, and there is no sense in which CPT can be thought of as a
discrete gauge symmetry.

In fact Hawking has argued that Black Holes can introduce a more
fundamental departure of the standard paradigm.\cite{Hawking:1976ra}
He proposed a generalization of quantum mechanics that allows pure
states to evolve into mixed states. He did this to address problems,
like the information paradox, that arise when trying to merge quantum
mechanics and general relativity. Page showed that the proposal of
Hawking leads to violation of CPT, and similarly for other
generalizations of QM.\cite{Page:1982fk} Since QM is very well established,
in order to test it one must look for extremely small deviations from
its predictions. It is very useful to have some idea of where to look,
a testable extension of QM that properly reproduces all existing data
to within present precision and accuracy. In 1989 Weinberg proposed a
version of quantum mechanics where the algebra of observables, what we
usually describe as matrices in Heisenberg's quantum mechanics, as
still forming an algebra but with a product that is no longer
associative (and of course, just as in ordinary QM, non-commutative).
I will not pursue this further, partly because Weinberg's formulation
is designed to incorporate Galilean invariance, rather than special
relativity. There have been other attempts at this, one by Kibble
trying to incorporate Lorentz invariance, and I refer you to
Ref.~\cite{Weinberg:1989cm} for further details and references.

This is an ongoing quest. I will turn to prospects shortly. But
Already in 1976 an experiment was conducted to test the validity of
QM. It was prompted by Eberhard's 1972 phenomenological analysis of QM
violation in the $K^0-\bar K^0$ system.\cite{Eberhard:1972am} Of course, no
positive signal of violations of QM was reported.\cite{Carithers:1975ri} This
is perhaps an example of bad luck. Or bad intuition, perhaps. But
since then the verification of the validity of the SM of electroweak
and strong interactions, which is a local QFT, allows us to argue, in
retrospect that, that the violations are expected to be characterized
by parameters of order $m_K^2/m_{Planck}\sim10^{-19}~\text{GeV}$,
below the resolution of those experiments.

\begin{figure}[t!] 
    \centering
    \includegraphics[height=7cm]{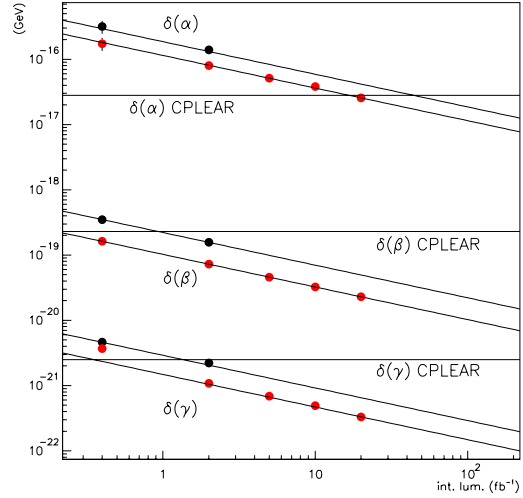}
    \caption{Limits obtained by CPLEAR\cite{Angelopoulos:2003hm}
and the limits that KLOE may be able to obtain on the CPT/QM violating
parameters $\alpha$, $\beta$ and $\gamma$ as a function of
luminosity. Two cases are shown. The black dots correspond to when
using the present inner tracker which has a vertex resolution of 0.9
while the red dots correspond to what would obtain if the inner
tracker had a resolution of 0.25. Figure taken from Ref.~\cite{Venanzoni:2010ix}}
    \label{fig:cplear}
\end{figure}

I do not intend to review the theory of violations to QM and CPT in
general, or even in particular for the neutral $K$ system. But I do
want to flash/show a few equations so you get an impression of where
modifications  to the ``normal'' case arise, that is to the CPT conserving
quantum mechanics.\footnote{For derivations and more complete
  discussion see Refs.~\cite{Maiani:1992ya,Ellis:1992dz,Huet:1994kr}} The short and long
eigenstates are defined familiarly:
\begin{equation*}
\begin{aligned}
|K_S\rangle &\propto (1+\epsilon_S)|K^0\rangle+(1-\epsilon_S)|\bar K^0\rangle \\
|K_L\rangle &\propto (1+\epsilon_L)|K^0\rangle+(1-\epsilon_L)|\bar
K^0\rangle
\end{aligned}
\qquad
\begin{aligned}
\epsilon_S=\epsilon+\Delta\\
\epsilon_L=\epsilon-\Delta
\end{aligned}
\qquad
\begin{aligned}
m_S-\textstyle{\frac{i}2}\Gamma_S=\bar m -\textstyle{\frac{i}2}\bar\Gamma-d
\\
m_L-\textstyle{\frac{i}2}\Gamma_L=\bar m
-\textstyle{\frac{i}2}\bar\Gamma+d
\end{aligned}
\qquad
d=\Delta m-\textstyle{\frac{i}2}\Delta\Gamma
\end{equation*}
The parameter $\epsilon$ is CP odd, CPT even, while $\Delta$ is CP
even but CPT odd. The masses and widths are given in terms of averages
$\bar m$ and $\bar \Gamma$ and a deviation $d$ with positive real and
imaginary parts.  The time development in the exclusive charged mode
decay 
\begin{equation}
R_{+-}(\tau)=\frac{N(K(\tau)\to\pi^+\pi^-)}{N(K(\tau=0)\to\pi^+\pi^-)}
\end{equation}
and  in the semileptonic decay,
\begin{equation}
\delta(\tau)=\frac{N(K(\tau)\to\pi^-\ell^+\nu)-N(K(\tau)\to\pi^+\ell^-\bar\nu)}{N(K(\tau)\to\pi^-\ell^+\nu)+N(K(\tau)\to\pi^+\ell^-\bar\nu)}\,,
\end{equation}
two extensively studied quantities  that the FPCP
audience know well, are given by:
\begin{align}
\delta(\tau)&=\frac{2\cos(\Delta m\tau)e^{-(\bar\Gamma+\alpha-\gamma)\tau}+2\text{Re}\;\epsilon_S^-e^{-\Gamma_S\tau}+2\text{Re}\;\epsilon_L^+e^{-\Gamma_L\tau}}{e^{-\Gamma_S\tau}+e^{-\Gamma_L\tau}}
\\
R_{+-}(\tau)&=e^{-\Gamma_S\tau}+R_Le^{-\Gamma_L\tau}+2|\bar\eta_{+-}|\cos(\Delta m\;\tau+\phi_{+-})e^{-(\bar\Gamma+\alpha-\gamma)\tau} 
\end{align}
For pure $K_L$ beam, 
\begin{equation}
\delta_L=2\text{Re}\,\epsilon^+_L\qquad\text{and}\qquad
R_L=|\epsilon^-_L|^2+\frac{\gamma}{\Delta\Gamma}+4\frac{\beta}{\Delta\Gamma}\,\text{Im}\,\left(\frac{\epsilon^-_Ld}{d^*}\right)\,,
\end{equation}
where $\epsilon_{L,S}^\pm=\epsilon_{L,S}\pm\frac{\beta}{d} $. The
formalism of Refs.~\cite{Ellis:1992dz,Huet:1994kr} has three
parameters,\footnote{Not to be confused with $\phi_{1,2,3}$ of the
  unitarity triangle!} $\alpha, \beta$ and $\gamma$, in addition to
the ones in the standard, CPT conserving, normal-QM one. These
equations display how they enter into these measurable quantities.
Setting these new parameters to zero reduce these expressions to the
standard ones. Note that they have units of mass. As mentioned above,
if the violations to QM or CPT arise from QG their natural size is
$10^{-19}$~GeV.

Is it crazy to go after this?  The plot in Fig.~\ref{fig:cplear},
taken from \cite{Venanzoni:2010ix}, shows the limits obtained by CPLEAR\cite{Angelopoulos:2003hm}
and the limits that KLOE may be able to obtain as a function of
luminosity.  As you can see KLOE cannot improve
on the CPLEAR bound on $\alpha$, which is more than two orders of
magnitude above where effects may be expected to show up. However,
KLOE can improve the bounds on $\beta$ and $\gamma$. Note that the
bound on $\gamma$ is already quite stringent. But it may be worth
pushing it: one can easily imagine that the order of magnitude
estimate of $10^{-19}$~GeV should be modified by a small coupling
constant;  after all, the GUT fine structure constant is about $1/40$.

\section{Violations to Lorentz Invariance}
Establishing the validity of Lorentz invariance (LI) has a long
history, going back to the famous Michelson-Morely experiment and many
others by their contemporaries. As in the case of departures from
Quantum Mechanics one starts by inventing a framework which
parametrizes the deviations by a small parameter that one can then
bound experimentally (for a review see \cite{Mattingly:2005re}).  But as
opposed to the QM case, it is very easy to invent a framework to
parametrize deviations from LI: simply take the SM and add terms to
the Lagrangian constructed of the same fields but that are not Lorentz
invariant. This is conveniently done by introducing would-be tensors
and writing Lorentz invariant terms that include these tensors, only
we take the tensors to be (coupling) constants,
\cite{Kostelecky:2002hh} {\it e.g.},
\begin{equation}
\label{noLI}
{\cal
  L}=-\textstyle{\frac14}F^{\mu\nu}F_{\mu\nu}-\textstyle{\frac14}(k_F)_{\mu\nu\lambda\sigma}F^{\mu\nu}F^{\lambda\sigma}\,.
\end{equation}
The theory in (\ref{noLI}) is equivalent to propagation in anisotropic
media: the relation between electric displacement, magnetic intensity
and electric-magnetic fields differs form that of in-vacuum:
\begin{equation}
\begin{pmatrix}\vec D\\ \vec H\end{pmatrix} = 
\begin{pmatrix}1+\kappa_{DE}&\kappa_{DB}\\ \kappa_{HE}& 1+\kappa_{HB}\end{pmatrix}
\begin{pmatrix}\vec E\\ \vec B\end{pmatrix}
\qquad\text{where}\qquad
\begin{aligned}
\kappa_{DE}^{jk}&=-2k_F^{0j0k}\,,\\
\kappa_{HB}^{jk}&={\textstyle\frac12}\epsilon^{jpq}\epsilon^{krs}k_F^{pqrs}\,,\\
\kappa_{DB}^{jk}&=-\kappa_{HE}^{kj}=\epsilon^{kpq}k_F^{0jpq}\,.
\end{aligned}
\end{equation}
It is convenient to define combinations of these $\kappa$'s that have
definite parity and that are either boost invariant or first order in
the frame velocity. Here are two of them, the only two that do not
produce birefringence: the parity even, boost independent
$\tilde\kappa_{e-}=\frac12(\kappa_{DE}-\kappa_{HB})-\frac13\text{Tr}\kappa_{DE}$,
and the parity odd, boost dependent,
$\tilde\kappa_{o+}=\frac12(\kappa_{DB}+\kappa_{HE})$. Astrophysical
measurements set bounds on the coefficients I have not written here,
through absence of birefringence, at some ridiculously low level,
$10^{-32}$ or so. The rest can be bound in laboratory experiments.
The table on next page shows, in units of $10^{-17}$, the results of
an experiment published earlier this year by a group from Berlin and
Bremen.\cite{Herrmann:2010ki}
\begin{table}
\begin{center}
\begin{tabular}{c|cc}
\hline\hline \rule{0pt}{12pt}
& Ref.~\cite{Herrmann:2010ki} & Ref.~\cite{Stanwix:2006jb} \\ \hline
$\kappa_{e-}^{XY}$ & -0.31 $\pm$ 0.73 & 29 $\pm$ 23 \\
$\kappa_{e-}^{XZ}$ & 0.54 $\pm$ 0.70 & -69 $\pm$ 22 \\
$\kappa_{e-}^{YZ}$ & -0.97 $\pm$ 0.74 & 21 $\pm$ 21 \\
$\kappa_{e-}^{XX} - \kappa_{e-}^{YY}$ & 0.80 $\pm$ 1.27 & -50 $\pm$ 47 \\
$\kappa_{e-}^{ZZ}$ & -0.04 $\pm$ 1.73 & 1430 $\pm$ 1790 \\
$\beta_{\oplus}\kappa_{o+}^{XY}$  &-0.14 $\pm$ 0.78 & -9 $\pm$ 26 \\
$\beta_{\oplus}\kappa_{o+}^{XZ}$ & -0.45 $\pm$ 0.62 & -44 $\pm$ 25 \\
$\beta_{\oplus}\kappa_{o+}^{YZ}$ &  -0.34 $\pm$ 0.61 & - 32 $\pm$ 23 \\
\hline\hline
\end{tabular}
\end{center}
\end{table}
The frequencies of two lasers, each stabilized
to one of two orthogonal cavities, are compared during active rotation
of their setup.  The factor $\beta_\oplus$, defined to be $10^{-4}$, accounts
for the Earth's orbital boost. A four year old result by Stanwix {\it
  et al} \cite{Stanwix:2006jb} is displayed side by side with this result, of
earlier this year, to highlight the pace of progress in the field
(also, a comparable result by yet another group was published last
year\cite{Eisele:2009zz}).  

The question arises as to what is the expected level of violation? And
why would some of the coefficients, the ones that can produce
birefringence, automatically vanish or are much suppressed (else one
would expect a priori equal order of magnitude and the search should
aim at improving the $10^{-32}$ bound that already exists for the
small kappas). These questions cannot be addressed by SME which is not a
theory, not even a model, but at best  only a parametrization.

\subsection{Scale of Lorentz violation? (Origin of Lorentz Violation?)}
It is natural to point to quantum gravity as responsible for
violations of Lorentz invariance. This is what proponents of doubly
special relativity (DSP) suggest. At least one of the proposals of DSP
argues that loop quantum gravity will result in an invariant
energy. As far as I know there is no complete argument that
demonstrates this.  This is not going to deter us, given that we
decided at the outset to look at non-mainstream ideas. But I should
point out that some independent proposals of the same idea have made
no attempt to connect this to loop QG.  Non-commutative spacetimes
have also been studied extensively. These, I think, are on a stronger
footing than DSP since for them one can consistently formulate a theory of
fields. There is no reason why the length scale should
be taken as the Planck Length, but since the theory originated from
work in string theory and there the fundamental scale is almost always
taken to be Planck's, the prejudice made it into non-commutative
spacetime.  And there are many others (Rainbow (energy dependent)
metric, $\kappa$-Minkowski, Hopf-algebras, spacetime foam, etc), which
I have no time to review (and besides, I know next to nothing about
them).  But they all have one thing in common, they modify the
dispersion relation of special relativity. They do not agree, however,
on what the modification ought to be. In fact, in most cases the
precise form of the modification is not known.  Take DSP. There is an
infinite class of dispersion relations that work. To see this
construct a non-linear realization of the Lorentz group as
follows. Take $F: P\to\mathcal{P}$ to be an invertible function from
the space of physical 4-momentum $P$ to a fictitious space
$\mathcal{P}$ of four vectors on which the Lorentz group acts
linearly. Then Lorentz transformations on $P$ are given by mapping to
the unphysical space, Lorentz transforming and then mapping back.
Now, since a Lorentz transformation leaves $\pi=0$ or $\pi=\infty$
invariant, we can choose a function that maps some particular value of
$p^0$ to either $\pi=0$ or $\pi =\infty$. As you can see this is very
general. Since we want to probe for a small effect we take any one of
these proposals and expand in powers of energy divided by
$M_{Planck}$. This is good enough for phenomenological studies and is
as much as we can expect from the state of theory. Note also that we
are only dealing here with modification to the dispersion
relation. This is as much as DSP can give you now. Full fledged
theories, like Non-commutative space time can also give you
modifications to the SM Lagrangian, and more. We content ourselves
with this for now and see where we get.

For ultra-relativistic particles a parametrization of
Ref.~\cite{AmelinoCamelia:2009pg} is as follows:
\begin{equation}
E\approx p+\frac{m^2}{p}-\frac{E}{\kappa}
\end{equation}
There is a lot of slop here. You can construct a non-linear
representation so that the correction only comes in at some higher
power of $E$. The sign and magnitude of the coefficient of the
correction are not fixed either. But this is a good starting point for
phenomenological tests.  A non-linear dispersion relation, applied to
massless particles, say, photons, gives an energy dependent speed of
light. This gives an immediate avenue for testing these ideas, looking
for energy dependent variations in time of travel over a fixed
distance, $\Delta t \approx (\Delta E/\kappa)L$. A source of photons
that is both very distant and has a wide (energy) spectrum is
required. Gamma Ray Bursts fit the bill. We do not understand these
sources well enough to argue that photons leave simultaneously.  But
we can be conservative and use the observed time delay as an upper
bound of the time delay produced by DSR and hence a lower bound $
\kappa > 1.3\times 10^{18}GeV\approx 0.10\,
M_{\text{Planck}}$,\cite{AmelinoCamelia:2009pg} which, remarkably, is close to the
Planck scale.

There are severe difficulties in constructing a quantum theory of
fields that respects invariance of a fundamental speed and of a
fundamental length. I do not wish to go into these difficulties. For
one thing I don't really understand the issues. Coraddu and Mignemi
suggest one can get glimpses of what such theory may be like by
considering a sort of first quantized version, a Schrodinger like
equation based on a modified dispersion relation.\cite{Coraddu:2009sb} This
is just like the Klein-Gordon theory but for a more complicated
energy-momentum dispersion relation. They take a particular version of
double-special relativity, get the following dispersion relation
\begin{equation}
  E=\frac{-\frac{m^2c^4}{\kappa}\pm\sqrt{\left(1-\frac{m^2c^4}{\kappa^2}\right)
      c^2\vec p^2+m^2c^4 }}{1-\frac{m^2c^4}{\kappa^2}}
\end{equation}
and then solve the Klein-Gordon-like equation. From the
non-relativistic limit they find that the inertial mass is not the
same as the invariant mass parameter,
\begin{equation}
m^\pm=\pm\frac{m}{1\pm \frac{mc^2}{\kappa}}\,.
\end{equation}
The minus sign is for a solution that is interpreted as a hole, so
$-m^-$ is the actual mass of the antiparticle.  While I do not
understand any of this would-be-formalism, I can certainly plug in
numbers. The limit on the $K^0-\bar K^0$ mass difference
is very good\cite{Amsler:2008zzb} so we can obtain 
\begin{equation}
  \kappa > \frac{2m c^2 }{(\Delta m/m)_{\text{max, exp}}}\approx 1.1\times 10^{18}~\text{GeV}
\end{equation}
The result is tantalizingly close to (but better than) the bound
obtained from time delays in GRBs, and suggests another
experimental direction in which  to push.

\section{Acausality and Nonlocality}
We have to be very careful when discussing violations to
causality. First of all, we must not mean that things can happen
without having a cause. Metaphysically we want to insist that every
effect has a cause, else we can replace religion for
science. Contemporary physics's view of causality has taken shape by
incorporating two additional principles: Lorentz covariance and
locality.  The causal-light-cone construction is based on this. That a
point can only influence another point if it is in its future
light-cone assumes that information from the point propagated no
faster than the speed of light and it acted on the second point at its
location.  Suppose we drop the assumption of locality and allow a mild
departure, whereby an interaction can be non-local but only at small,
microscopic range (this can be phrased in a Lorentz invariant
fashion). Then we can send a signal that in the future influences an
object outside our light-cone. Now, as we learn in elementary courses
on special relativity, superluminal communication may lead to
paradoxes. For example, the grandfather paradox says that if you can
shoot a bullet that travels faster than the speed of light then there
is a frame where it travels back in time; you can shoot your
grandfather before your mom was conceived. One is tempted to conclude
that non-local interactions, even if Lorentz invariant, are ruled out
by consistency (that is, by requiring no grandfather paradoxes).  I
believe that would be rushing to conclusions. It is not trivial to connect
even in principle a sequence of non-local interactions to produce
effectively superluminal information propagation.  Furthermore, one
has to incorporate quantum mechanics into the picture. 

Schrodinger's equation is perfectly causal: given an initial
wavefunction it gives a method for computing evolution of the
wavefunction into a later time. Of course, we want to incorporate into
this Lorentz covariance, so we are really talking about QFT. And while
we do in principle have a Schrodinger equation for the field functional
in QFT, the issue of causality is a murky one. If you open a textbook
on QFT you may or may not find a discussion of causality, other than
the common requirement that commutators vanish outside the
light-cone. But it is easy to find acausal theories with such
commutators. So this is not a sufficient condition. There are also
conditions on analyticity of amplitudes in the upper half of the
complex energy plane. But these conditions are too strong.  It seems
that Schrodinger evolution plus Lorentz covariance should be a recipe
for causal QFT, but this must fail if we have non-locality on
arbitrarily long distances. 

So as confusion reigns the one thing limited minds like mine can turn
to is explicit examples. It is straightforward to invent such a
theory. Just add to any normal looking theory (say the SM of EW
interactions) some additional terms with higher derivatives. The new
theory is still renormalizable, Lorentz invariant and has controlled
non-locality, the distance scale of the non-locality being fixed by
the dimensional parameter $\ell$ that accompanies the higher
derivatives. But this is tough business: there are many publications
on theories that are hopelessly sick because non-localities can
introduce ghosts (states with negative metric) or bad instabilities or
both. As far as I know the Lee-Wick procedure is the best shot at a
consistent quantization with indefinite metric.\cite{Lee:1969fy}

In theories of this sort you have some unstable states, resonances,
that behave weirdly. They are very heavy, their mass of order of the
inverse of the scale of non-locality. They may or may not be
narrow. It is amusing to consider the case of a narrow resonance of
this type, which I will call a ``Lee-Wick'' particle.  To understand
better let me remind you of the behavior of a resonance produced in a
fixed target experiment. A particle beam impinging on a fixed target
produces a signal consisting of, {\it e.g.,} pair production behind
the target. The location of the production of the pair is determined
from extrapolation back from the detectors, and the rate of production
of these pairs decays exponentially away from the target. This is
interpreted as a resonance being produced followed by its decay in
flight. The number of produced pairs falls off exponentially with
distance from the target, which is related to the life-time (or
inverse width) of the resonance times its velocity. 

Now for the Lee-Wick case:\cite{Coleman:1969xz} The signal would be
exactly the same except that the pair is produced ahead of the target,
as if the resonance decays before the interaction takes place, so it
appears it travels backwards in time. The number of pairs produced
again decreases exponentially with distance form the target. You can
imagine trying to construct a grandfather's-like paradox in this
situation: to a particle detector connect a mechanism that quickly
removes the target, so that one detects the collision products but
there is no target for the collision to take place. But this does not
happen, if you remove the target the collision does not occur.  If the
distance scale $\ell$ is microscopic, say a TeV$^{-1}$ or smaller,
this effect is tremendously difficult to observe (and may explain why
this sort of acausality has not been seen). A TeV is the relevant
scale if the higher derivatives are the cure of the hierarchy problem.\cite{Grinstein:2007mp}

However there may be better ways to test for these effects. It is
still not easy, but probably doable.  The idea is to study this
resonance's phase shift. While the shape of the LW resonance itself is
very normal, the phase shift quickly {\it decreases} across the
resonance (it rotates {\it clockwise} in the Argrand diagram)!  To
measure a phase shift one needs to study exclusive processes. So I do
not think this is LHC physics.  Instead, once the LW resonances are
discovered at the LHC we will need the ILC to verify their unusual
clockwise phase shifts. But don't hold your breath waiting for a time
machine out of this!

\section{Final Remarks}
As I remarked earlier, particle physicists have done a remarkable job
of establishing the validity of the SM at a great level of
precision. I also reminded us why we believe it is incomplete: I do
believe exciting discoveries are about to occur at the LHC. We need an
explanation for the hierarchy problem, for neutrino masses, for
baryogenesis, for dark matter and energy. We would really like to have
an answer to ``who order that!'' and to be able to calculate all
masses and mixing angles in quark and lepton sectors.  And is there
anything to the apparent unification in the MSSM? Some answers to
these are bound to reveal themselves and the process of discovery will
be fun and exciting.  But things can be even wilder. And we have to be
open to those possibilities if we are to find out. I argued that it
was along these lines that FP and CP were born. So it makes sense to
imagine ``what if.'' Now, not a single idea I presented is
solid. Worse, much of the theory I presented is flimsy, or not well
motivated, or both. Some of it may not even be mathematically or
metaphysically consistent. But it may give us the lamppost with the
light to look under.  Perhaps in some years the FPCP in the title of
the conference will refer to something else. To something we ought to
continue doing, looking for departures from the standard
paradigm which may look very different from our current SM. Something like 
``Future and Pseudo-Causal Physics...''

\end{document}